# A Study of Complete and Incomplete Reactions of $^{12}$C +$^{169}$Tm System At Energy Range ≈ 4.16 –7.5 Mev/Nucleon

Getahun Kebede  (author)

Department of physics,

Hawassa University

Tel: +251912212883

E-mail: getahunkebede@mtu.edu.et

## Abstract

In this paper an attempt was to measure, the excitation functions of $^{169}$Tm($^{12}$C, 4n)$^{177}$Re, $^{169}$Tm($^{12}$C, 5n)$^{176}$Re, $^{169}$Tm($^{12}$C, $\alpha$n)$^{176}$Ta , 169 Tm($^{12}$C, $\alpha$2n)$^{175}$Ta , $^{169}$Tm($^{12}$C, $\alpha$3n)$^{174}$Ta , $^{169}$Tm($^{12}$C, $\alpha$4n)$^{173}$Ta  and $^{169}$Tm($^{12}$C, 2$\alpha$2n)$^{171}$Lu reaction channels populated in the interaction of $^{12}$C projectile with $^{169}$Tm target were considered in order to investigate the mechanisms of complete and incomplete fusion reactions. The theoretically predicted excitation functions using PACE4 code were compared with the previously measured excitation functions. For non–α emitting channels cross-section values predicted by PACE 4 in general were found to be in good agreement with the experimentally measured values. However, for α-emitting channels the measured cross-section values were found to be higher than the values predicted by PACE4. The observed disagreement may be credited to projectile break-up in the vicinity of n-n interaction.

# Introduction

Most nuclear reactions are studied by inducing a collision between two nuclei (nucleon- nucleon reaction) where one of the reacting nuclei is at rest (the target nucleus) while the other nucleus (the projectile nucleus) is in motion. Projectiles heavier than α-particle (i.e A≥4) are commonly regarded as heavy ions and become used for bombarding the target nuclei.

It is now generally recognized that several reaction mechanisms are operative in heavy ion-induced reactions below 10MeV/nucleon. In fact the cluster structure has been suggested as one of the factors leading to forward peaked α-particles in ICF reactions. While CF has been defined as the capture of total charge or mass of the incident projectile by the target nucleus.

However, the first evidence of ICF reactions was presented by Kauffmann and Wolfgang [1], by studying $^{12}$C+ $^{102}$Rh system at energy range of 7-10 MeV/nucleon, where strongly forward peaked angular distributions of light-nuclear-particles were observed. Britt and Quinton [2], found similar observations in the $^{16}$O+ $^{209}$Bi reactions at energies range 7-10 MeV/nucleon. In these measurements, significantly large yield of direct α-particles of mean energy roughly corresponding to the projectile velocity at the forward cone has been observed [3-7].

Meanwhile, the IFC system (reduced CN) forms with relatively less mass/charge and excitation energy (due to partial fusion of projectile), but at high angular-momenta (imparted due to non-central interactions) as compared to the CN formed via CF.

In the past various studies were done on the mechanism of CF and ICF reactions. Recently Amanuel et al [8] studied the role of break up process in the fusion of the $^{12}$C + $^{52}$Cr system at several beam energies from ≈ 4-7MeV/nucleon. It was found that from non-α-emitting channels the experimentally measured excitation functions were, in general found to be in good agreement with PACE4 predicted. However, for α-emitting channels the measure EFs were higher than PACE4 predicted which is attributable for ICF reactions.

A number of studies in the past were confined to beam energies greater than 10 MeV/nucleon and the reaction mechanism have been reasonably explained by the available models. Various dynamical models, such as, Sum rule model [9],break-up fusion (BUF) model [10] and promptly emitted particle model [11] have been proposed to explain the mechanism of ICF reactions. However, no theoretical model is available so far fully to explain the gross features of experimental data available below E/A=10 MeV/nucleon. Despite a number of attempts in the past none of the available models are able to reproduce the experimental data obtained at energies as low as ≈ 4-8Mev/nucleon. There is no fully investigation is conducted on ICF processes, that is why still needs further study, especially at relatively low bombarding energy 10MeV/nucleon since a clear systematic study and complied data are available for only a few projectile target systems.

In this work the experimentally measured (EXFOR data) EFs for reactions $^{169}$Tm($^{12}$C, 4n)$^{177}$Re, $^{169}$Tm($^{12}$C, 5n)$^{176}$Re, $^{169}$Tm($^{12}$C, αn)$^{176}$Ta, $^{169}$Tm($^{12}$C, α2n)$^{175}$Ta, $^{169}$Tm($^{12}$C, α3n)$^{174}$Ta, $^{169}$Tm($^{12}$C, α4n)$^{173}$Ta, and $^{169}$Tm($^{12}$C, 2α2n)$^{171}$Lu in the incident energy range 50 - 90MeV were compared with theoretical predictions based on PACE4 codes. For predication of the measured excitation function the theoretical model of PACE 4 was used with 100000 cascades.

## 2. COMPUTER CODE AND FORMULATION

There are various computer codes such as PACE4, CASCADE, COMPLETE CODE (modified of ALIC- 91) which are available to perform such statistical model calculations. The PACE4 [12] code was chosen to be used in the present work since it is easily available and proved to be one of the most reliable and promising theoretical model for the compound nuclear reactions. And analysis with computer code PACE4 within the consideration of Hauser-Feshbach formulation also discussed in this section. The code uses the BASS model for CF cross section calculation and used as Monte Carlo procedure to determine the decay of sequence of an excited nucleus using Hauser-Feshbach formalism.

In this statistical code for neutrons, protons and α-particles the default optical model parameters are used. In addition code has been modified to take into account the excitation energy dependence of the level density parameter using the prescription Kataria et al [13]. It should be pointed out that the ICF and PE-emission are not taken into consideration in this code. The process of de-excitation of the excited nuclei was calculated using code PACE4 which follows the correct procedure for angular momentum coupling at each stage of de-excitation.

As a result to compare the measured EF's with theoretical predication obtained from PACE4 for possible residues populated in reaction. Cross-section is deduced using Morgenstern et al [14].

$$\sum \sigma_{CF}^{theo} = \sum \sigma_{non-\alpha\ emit}^{exp} + \sum \sigma_{\alpha\ emit}^{theo} \qquad 1$$

In order to extract more information regarding how ICF contributes to total fusion reaction cross-section is given by:

$$\sigma_{TF} = \sum \sigma_{CF}^{theo} + \sum \sigma_{ICF} \qquad 2$$

From this cross-section the total ICF cross-section can be found using an expression of

$$\sum \sigma_{ICF} = \sigma_{TF} - \sum \sigma_{CF}^{theo} \qquad 3$$

The enhancement from the theoretical predictions points towards the presence of ICF process in the formation of all ERs, the contribution of ICF in the formation of all $\alpha$-emitting channels has been calculated as

$$\sum \sigma_{ICF} = \sum \sigma_{\alpha\,emit}^{exp} - \sum \sigma_{\alpha\,emit}^{theo} \qquad 4$$

The contribution of ICF in the formation of all $non-\alpha$-emitting channels has not been observed due to no $\alpha$ cluster is populated by break up process.

$$\sum \sigma_{ICF} = \sum \sigma_{non-\alpha\,emit}^{exp} - \sum \sigma_{non-\alpha\,emit}^{theo}, \quad \text{but for non } \alpha \text{ emitting channel } \sum \sigma_{ICF} = 0$$

i.e $\quad \sum \sigma_{non-\alpha\,emit}^{exp} = \sum \sigma_{non-\alpha\,emit}^{theo} \Rightarrow \sigma_{non-\alpha\,emit}^{exp} = \sigma_{non-\alpha\,emit}^{theo}$, and it is true for each individual ERs.

## 3. RESULT AND DISCUSSIONS

In this work the excitation functions for seven residues produced in the $^{12}C + ^{169}Tm$ system were studied. The experimentally measured excitation functions were compared with the theoretical predictions obtained from the code PACE4. The experimental cross-section and energy are obtained from IAEA data source (EXFOR) Library [15].

In order to show the effect of variation of $K$ on calculated EFs, different values of K= 8, 10, 12 and 14 have been tested, and are shown in Fig.3.1 (a). Therefore in this work, a value of $K = 8$ is found to give a satisfactory reproduction of experimental data for CF-channels within the experimental uncertainties and have been chosen confidentially for other $\alpha$-emitting channels.

3.1. **Evaporation Residues Populated Through Non-$\alpha$-Emitting (12C, xn) Channels**
   A) ($^{12}C$, 4n) channel

For the representative ($^{12}C$, 4n) channel values of the level density parameter K (K=8, 10, 12 and 14) were varied to fit the experimental data and the results are displayed in Fig.3.1. The $^{177}Re$ residue was produced when $^{12}C$ projectile completely fused with $^{169}Tm$ target lead to the formation of excited compound nucleus $^{181}Re^*$. The excited CN, $^{181}Re^*$, decay through the emission of four neutrons that leads to the formation of isotope $^{177}Re$. In reaction equation form, it is written as:

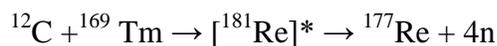
$$^{12}C + ^{169}Tm \rightarrow [^{181}Re]^* \rightarrow ^{177}Re + 4n$$

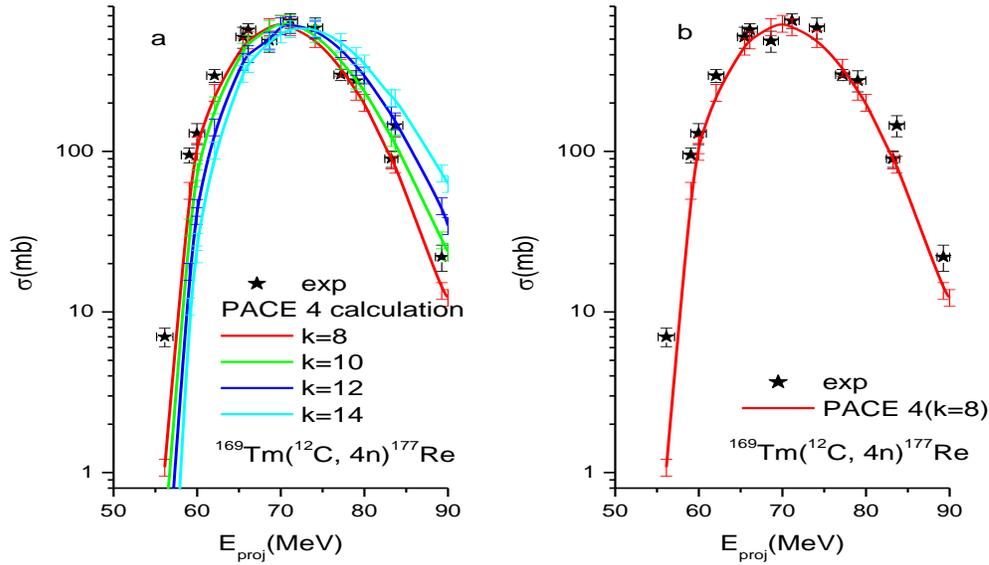

Fig.3.1. Experimentally Excitation function for the $^{169}$Tm($^{12}$C, 4n) $^{177}$Re reaction used for studying the effect of the value of k on theoretically calculated results expected to be populated by CF compared with their theoretical prediction (PACE4) a) at different k value and b) at k=8 that has been best fitted within the energy range ≈ 4.16–7.5 MeV/nucleon.

As can see from Fig.3.1 the theoretically calculated excitation function corresponding to the level density parameter K=8 in general satisfactorily reproduced the experimentally measured EFs for residue $^{177}$Re produced in the CF of $^{12}$C projectile with $^{169}$Tm target. In the present calculation a value of K=8 will be used for all other residues populated in $^{12}$C + $^{169}$Tm system. Further it may be mentioned that the general trends and shape of the measured EFs for the CF residues populated 4n channels are satisfactorily reproduced by PACE4 calculations with uncertainties for entire energy region as shown in Fig.3.1.

### B) ($^{12}$C, 5n) channel

The $^{176}$Re residue was produced when $^{12}$C projectile completely fused with $^{169}$Tm target lead to the formation of excited compound nucleus $^{181}$Re$^*$. The excited CN, $^{181}$Re$^*$, decay through the emission of five neutrons that leads to the formation of isotope $^{176}$Re. In reaction equation form, it is written as:

$$^{12}C + ^{169}Tm \rightarrow [^{181}Re]^* \rightarrow ^{176}Re + 5n$$

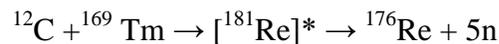

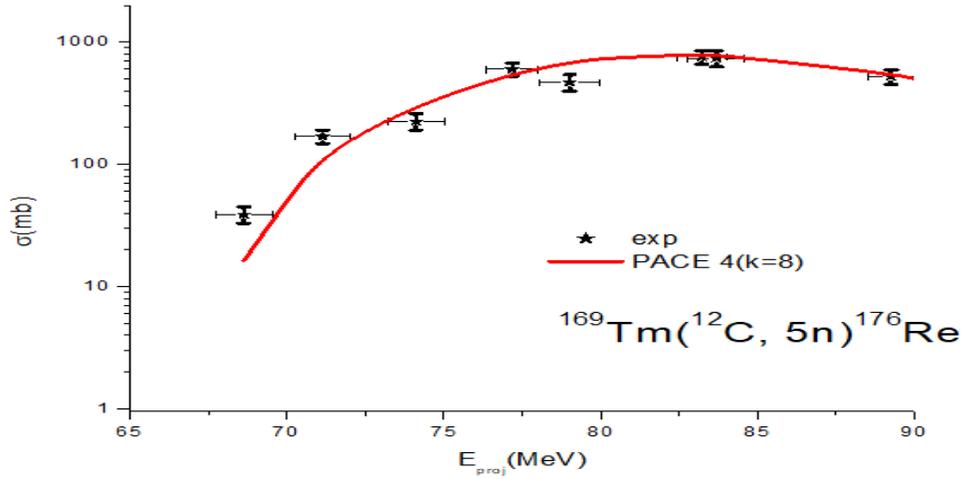

Fig.3.2. Experimentally Excitation function for the $^{169}$Tm($^{12}$C, 5n) $^{176}$Re reaction populated by CF compared with their theoretical prediction (PACE4) at k=8 within the energy range ≈ 4.16–7.5 MeV/nucleon.

The experimentally measured EFs along with theoretical predictions obtained using the PACE4 code residues populated via non α-emitting channels ($^{12}$C, 5n) is shown in Fig.3.2. The theoretically calculated excitation function corresponding to the level density parameter K=8 in general satisfactorily reproduced the experimentally measured EFs for residue $^{176}$Re produced via the CF of $^{12}$C projectile with $^{169}$Tm target.

### 3.2 Evaporation Residues Populated Through α- emitting ($^{12}$C, αxn) channels

#### A) ($^{12}$C, αn) channel

The $^{176}$Ta residue was produced when $^{12}$C projectile completely fused with $^{169}$Tm target lead to the formation of excited compound nucleus $^{181}$Re$^*$ and $^{12}$C incompletely fused with $^{169}$Tm lead to the formation of composite system $^{177}$Ta. This residue may be formed via CF and/or ICF in interaction of $^{12}$C with $^{169}$Tm following two processes. i). In case of CF, the composite system $^{181}$Re$^*$, decay through the emission of one α cluster and one neutrons that leads to the formation of isotope $^{176}$Ta. ii) The same residue is formed by ICF of $^{12}$C breaks in to α+$^8$Be and $^8$Be fuses with the target leaving α as spectator to form an incompletely fused composite system [$^{177}$Ta]$^*$ may then decay via one neutrons (n).

In reaction equation form, it is written as:

I.  Complete fusion (CF) of $^{12}$C:

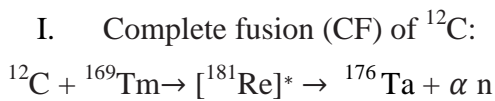

$^{12}$C + $^{169}$Tm → [$^{181}$Re]$^*$ → $^{176}$Ta + α n

Where α is as participant, not as spectator.

II. Incomplete fusion (ICF) of $^{12}$C

$$^{12}C(8Be + \alpha) + ^{169}Tm \rightarrow \alpha + [^{177}Ta]^* \rightarrow ^{176}Ta + \alpha + n$$

(α as a spectator which is not participate on the reaction).

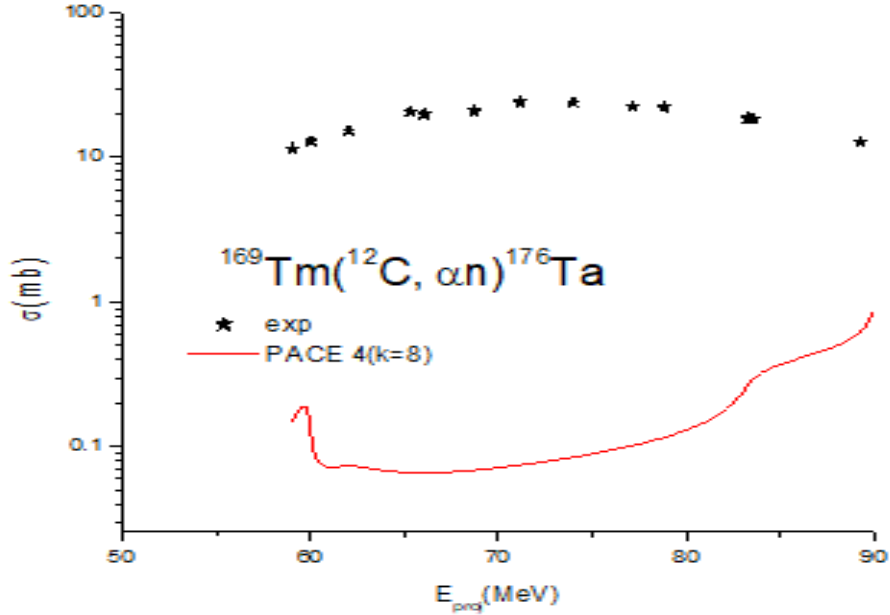

Fig.3.3. Experimentally Excitation function for the $^{169}$Tm($^{12}$C, αn)$^{176}$Ta reaction compared with their theoretical prediction (PACE 4).

As can be seen from Fig.3.3, the experimentally measured EFs are higher as compared to the theoretical predictions. Since the PACE 4 code doesn't take ICF in to account, therefore the enhancement in the experimentally measured cross sections are attribute to the contribution of ICF of $^{12}$C with $^{169}$Tm target.

B) ($^{12}$C, α2n) channel

The $^{175}$Ta residue was produced when $^{12}$C projectile completely fused with $^{169}$Tm target lead to the formation of excited compound nucleus $^{181}$Re$^*$ and $^{12}$C incompletely fused with $^{169}$Tm lead to the formation of composite system $^{177}$Ta. This residue may be formed via CF and/or ICF in interaction of $^{12}$C with $^{169}$Tm following two processes. i) In case of CF, the composite system $^{181}$Re$^*$, decay through the emission of one α cluster and two neutrons that leads to the formation of isotope $^{175}$Ta. ii) The same residue is formed by ICF of $^{12}$C breaks in to α+$^8$Be and $^8$Be fuses with the target leaving $\alpha$ as spectator to form an incompletely fused composite system $[^{177}Ta]^*$ may then decay via two neutrons (2n).

In reaction equation form, it is written as:

I   Complete fusion(CF) of $^{12}C$:

$$^{12}C + ^{169}Tm \rightarrow [^{181}Re]^* \rightarrow ^{175}Ta + \alpha 2n$$

($\alpha$ is as participant in the reaction, not as spectator)

II   Incomplete fusion(ICF) of $^{12}C$:

$$^{12}C(^8Be + \alpha) + ^{169}Tm \rightarrow \alpha + [^{177}Ta]^* \rightarrow ^{155}Ta + \alpha + 2n$$

($\alpha$ as a spectator, which is not participate on the reaction).

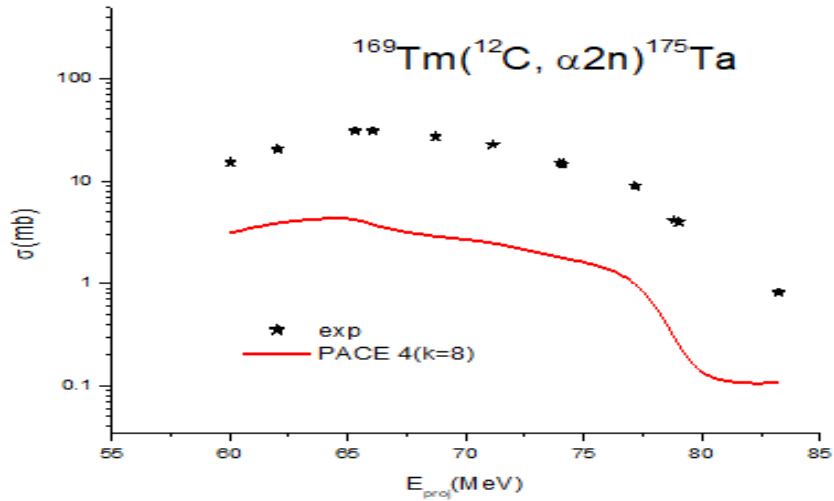

Fig.3.4. Experimentally Excitation function for the $^{169}Tm(^{12}C, \alpha 2n)^{175}Ta$ reaction compared with their theoretical prediction (PACE4).

As can be seen from Fig.3.4, the experimentally measured EFs are higher as compared to the theoretical predictions. As such, it may again be inferred that major contribution of the enhancement for the production of these residues comes from ICF processes, which are not considered in these calculations in the interaction of $^{12}C$ with $^{169}Tm$ target.

### C) ($^{12}C, \alpha 3n$) channel

The $^{174}Ta$ residue was produced when $^{12}C$ projectile completely fused with $^{169}Tm$ target lead to the formation of excited compound nucleus $^{181}Re^*$ and $^{12}C$ incompletely fused with $^{169}Tm$ lead to the formation of composite system $^{177}$Ta. This residue may be formed via CF and/or ICF in interaction of $^{12}$C with $^{169}$Tm following two processes. i) In case of CF, the composite system $^{181}$Re$^*$, decay through the emission of one $\alpha$ cluster and three neutrons that leads to the formation of isotope $^{174}$Ta. ii) The same residue is formed by ICF of $^{12}$C breaks in to $\alpha+^8$Be and $^8$Be fuses with the target leaving $\alpha$ as spectator to form an incompletely fused composite system [$^{177}$Ta]$^*$ may then decay via two neutrons (3n).

In reaction equation form, it is written as:

    I    Complete fusion (CF) of $^{12}$C:

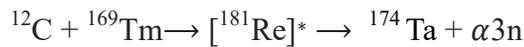

$$^{12}\text{C} + {}^{169}\text{Tm} \rightarrow [^{181}\text{Re}]^* \rightarrow {}^{174}\text{Ta} + \alpha 3n$$

($\alpha$ is as participant, not as spectator).

    II    Incomplete fusion(ICF) of $^{12}$C:

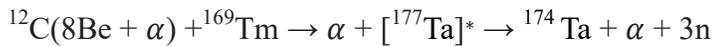

$$^{12}\text{C}(8\text{Be} + \alpha) + {}^{169}\text{Tm} \rightarrow \alpha + [^{177}\text{Ta}]^* \rightarrow {}^{174}\text{Ta} + \alpha + 3n$$

($\alpha$ as a spectator which is not participate on the reaction).

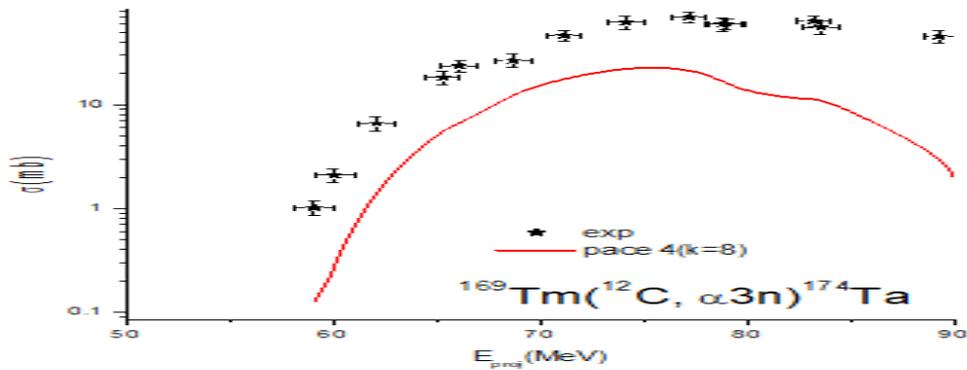

Fig.3.5. Experimentally Excitation function for the $^{169}$Tm($^{12}$C, $\alpha 3n$)$^{174}$Ta reaction compared with their theoretical prediction (PACE4).

The experimentally measured cross-section is relatively higher than the theoretical predictions as shown from Fig.3.5. Since the code PACE4 doesn't take ICF into account, therefore the enhancement in the experimentally measured cross-sections are attributable to the contributions of ICF of $^{12}$C with $^{169}$Tm target.

### D) ($^{12}$C, $\alpha 4n$ ) channel

The $^{173}$Ta residue was produced when $^{12}$C projectile completely fused with $^{169}$Tm target lead to the formation of excited compound nucleus $^{181}$Re$^*$ and $^{12}$C incompletely fused with $^{169}$Tm lead to the formation of composite system $^{177}$Ta. This residue may be formed via CF and/or ICF in interaction of $^{12}$C with $^{169}$Tm following two processes. i) In case of CF, the composite system $^{181}$Re$^*$, decay through the emission of one α cluster and four neutrons that leads to the formation of isotope $^{173}$Ta. ii) The same residue is formed by ICF of $^{12}$C breaks in to α+$^8$Be and $^8$Be fuses with the target leaving α as spectator to form an incompletely fused composite system [$^{177}$Ta]$^*$ may then decay via four neutrons (4n).

In reaction equation form, it is written as:

I. Complete fusion(CF) of $^{12}$C

$$^{12}C + {}^{169}Tm \rightarrow [^{181}Re]^* \rightarrow {}^{173}Ta + \alpha 4n$$

Where α is as participant not as spectator.

II. Incomplete fusion(ICF) of $^{12}$C

$$^{12}C(^8Be + \alpha) + {}^{169}Tm \rightarrow \alpha + [^{177}Ta]^* \rightarrow {}^{173}Ta + \alpha + 4n$$

α as a spectator which is not participate on the reaction( act as observer).

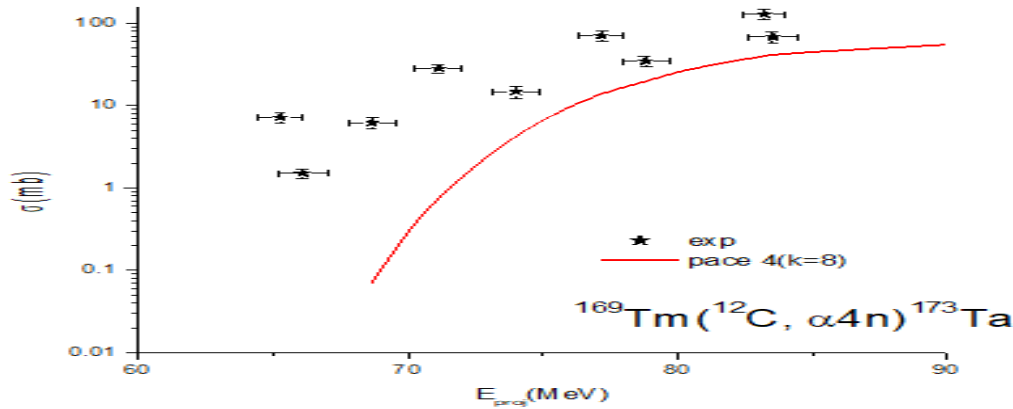

Fig.3.6. Experimentally Excitation function for the $^{169}$Tm($^{12}$C, α4n)$^{173}$Ta reaction compared with their theoretical prediction (PACE4).

The experimentally measured cross-section exhibit a significant enhancement compared to the theoretical predictions as can be seen from fig.3.6. As such, it may again be inferred that major contribution of this enhancement comes from ICF processes, which are not considered in these calculations.

### E) ($^{12}$C, $2\alpha 2n$) channel

The $^{171}$Lu residue was produced when $^{12}$C projectile completely fused with $^{169}$Tm target lead to the formation of excited compound nucleus $^{181}$Re* and $^{12}$C incompletely fused with $^{169}$Tm lead to the formation of composite system $^{173}$Lu. This residue may be formed via CF and/or ICF in interaction of $^{12}$C with $^{169}$Tm following two processes. i) In case of CF, the composite system $^{181}$Re*, decay through the emission of one $2\alpha$ cluster and two neutrons that leads to the formation of isotope $^{171}$Lu. ii) The same residue is formed by ICF of $^{12}$C breaks in to $^8$Be($\alpha + \alpha$) + $\alpha$ and $\alpha$ fuses with the target leaving $^8$Be as spectator to form an incompletely fused composite system $[^{173}$Lu$]^*$ may then decay via two neutrons (2n).

In reaction equation form, it is written as:

I. Complete fusion (CF) of $^{12}$C:

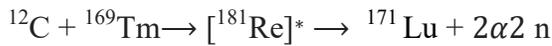

$$^{12}\text{C} + {}^{169}\text{Tm} \rightarrow [^{181}\text{Re}]^* \rightarrow {}^{171}\text{Lu} + 2\alpha 2\text{n}$$

($2\alpha$ is as participant in the reaction system, not as spectator).

II. Incomplete fusion (ICF) of $^{12}$C:

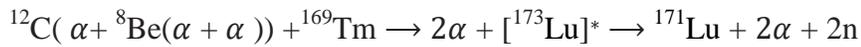

$$^{12}\text{C}(\alpha + {}^8\text{Be}(\alpha + \alpha)) + {}^{169}\text{Tm} \rightarrow 2\alpha + [^{173}\text{Lu}]^* \rightarrow {}^{171}\text{Lu} + 2\alpha + 2\text{n}$$

($2\alpha$ as a spectator which is not participate on the reaction)

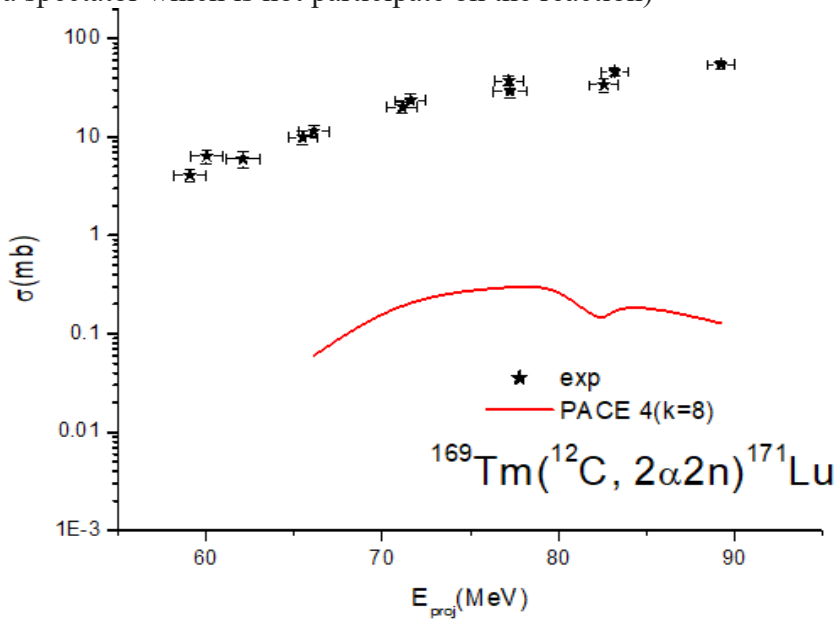

Fig.3.7. Experimentally Excitation function for the $^{169}$Tm($^{12}$C, $2\alpha 2n$)$^{171}$Lu reaction compared

with their theoretical prediction (PACE4).

In case of reaction $^{169}$Tm($^{12}$C, $2\alpha 2n$)$^{171}$Lu, as can been seen from Fig.3.7 the experimentally measured EF exceeds the theoretical EF, which again indicates that ICF plays an important role. Since, theoretical calculations of PACE 4 does not take into account the ICF, it may be inferred that a significant part of these reactions involving $2\alpha$ -emission channels go through ICF largely, at these energies.

Further it is obvious that α-emitting channels have contributions coming from ICF reactions. Fig.3.8.displayed the sum of experimentally measured cross section $\sum \sigma_\alpha(\exp)$, along with the sum of PACE4 cross section $\sum \sigma_\alpha(\text{Theo})$, As can be seen from this figure there is a clear gap between these two values which is attributable to the contribution coming from ICF reactions. Further from this figure the increasing separation between $\sum \sigma_\alpha(\exp)$ and $\sum \sigma_\alpha(\text{Theo})$ indicates that when projectile energy is increased the contribution of the ICF also relatively increased.

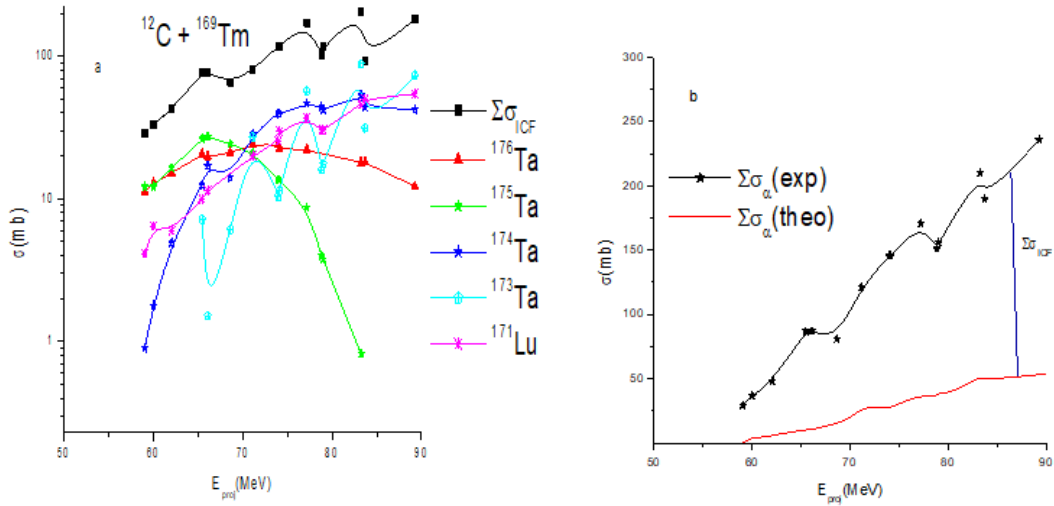

Fig.3.8. Sum of experimentally measured EFs of channels ($\sum \sigma_{\alpha xn+2\alpha 2n}^{exp}$) are compared with that predicted by statistical model code PACE4 ($\sum \sigma_{\alpha xn+2\alpha 2n}^{pace\ 4}$).

## 4. CONCLUSION

In this work, the excitation function of $^{176, 177}$Re, $^{173,174,175,176}$Ta and $^{171}$Lu evaporation residues produced via CF and/or ICF reactions in the interaction of $^{12}$C projectile with $^{169}$Tm target at energies ≈ 4.16 - 7.5MeV/nucleon were studied. The experimentally measured EFs were compared with theoretical calculations done using the PACE4 code. For non-α emitting channels the experimentally measured production cross-sections were found to be in good agreement with theoretical. In such reactions a case we expect the projectile is completely fused with the target, which is a mechanism that can be effectively described by PACE4. However for α emitting channel the theoretical predictions did not reproduce the experimental measured EFs. The observed enhancement may be attributed to the ICF processes from break-up of $^{12}$C projectile. $^{12}$C projectile breaks into $^{8}$Be and an- particle, and $^{8}$Be fragment fuses with $^{169}$Tm, forming the incompletely composite nucleus, followed by the emission of neutrons and α-particle. The present analysis showed that in heavy-ion induced reaction mechanisms study, the contribution from ICF is an important component of fusion reactions in particular at higher energy points. Furthermore, the present study showed ICF cross-section in general increases with increase in projectile energy. So it may be possible to conclude that complete and incomplete fusion reaction play important roles in heavy ion induced reaction mechanism studies.